\documentclass{aa}
\usepackage{graphicx}
\usepackage{txfonts}
\usepackage{times}
\usepackage{natbib}
\usepackage{color}

\def\eq#1{\begin{equation} #1 \end{equation}}
\def\E#1{\hbox{$10^{#1}$}}
\def\x      {\hbox{$\times$}}
\def\about  {\hbox{$\sim$}}
\def\sub#1{_{\rm #1}}

\def\Rin    {\hbox{$R\sub{in}$}}
\def\Rout   {\hbox{$R\sub{out}$}}
\def\Rc     {\hbox{$R\sub{c}$}}
\def\DR     {\hbox{$\Delta R$}}
\def\DH     {\hbox{$\Delta H$}}
\def\dv     {\hbox{$\delta v$}}
\def\Dv     {\hbox{$\Delta v$}}
\def\No     {\hbox{$N_0$}}
\def\Mo     {\hbox{M$_\odot$}}
\def\Lo     {\hbox{L$_\odot$}}
\def\kms    {\hbox{km\,s$^{-1}$}}
\def\deg    {\hbox{$^\circ$}}
\def\Nbar   {\hbox{$\bar{N}$}}
\def\Nmax   {\hbox{$N\sub{max}$}}

\def\sp     {\phantom{M}}
\def\CH#1   {{\large\bf #1}}

\begin{document}

   \title{A compact starburst ring traced by clumpy \\ OH megamaser emission}

   \author{R. Parra\inst{1}, J. E. Conway\inst{1} \and M. Elitzur \inst{2} \and
      Y. M. Pihlstr\"{o}m \inst{3} }

   \offprints{R. Parra\\ \email{rodrigo@oso.chalmers.se}}
   \institute{Onsala Space Observatory, S-43992 Onsala,Sweden
   \and Department of Physics and Astronomy, University of Kentucky,
   Lexington, KY 40506, USA
   \and Department of Astronomy, California Institute
   of Technology, Pasadena, CA  }

   \authorrunning{Parra et al.}

  
\abstract{

We model the OH megamaser emission from the luminous infrared galaxy IIIZw35 as
arising from a narrow rotating starburst ring of radius 22 pc enclosing a mass
of 7\x\E{6} \Mo. We show how both the compact and  apparently diffuse maser
emission from this ring can arise from a single phase of unsaturated maser
clouds amplifying background radio continuum. The masing clouds are estimated to
have a diameter of $<0.7$pc and internal velocity dispersion of \about 20\kms.
We find that the clouds are neither self-gravitating nor pressure confined but
are freely expanding. Their dispersal lifetimes may set the vertical thickness
of the ring. For an estimated internal density of $3\times10^{3}$ cm$^{-3}$,
cloud masses are of order 24\Mo. The observed spectral features and velocity
gradients indicate  that the clouds must be outflowing and escaping the nucleus.
The cloud mass outflow rate is  estimated to be 0.8\Mo yr$^{-1}$, while the star
formation rate is \about 19\Mo yr$^{-1}$. Associated ionised gas, possibly
generated from dissipated  clouds, provides free-free absorption along the
source axis, explaining the observed East-West asymmetries. We show that the
clumpiness of a maser medium can  have a dramatic effect on what is observed
even in a relatively low gain OH  megamaser. Specifically, in IIIZw35 our clumpy
maser model naturally explains the large line to continuum ratios, the large
1667MHz:1665MHz line ratios and the wide velocity dispersions seen in the
compact maser spots. Other astrophysical masers showing both compact and
apparently diffuse emission might be  explained by similar clumpy structures.

 }

\maketitle

\section{Introduction}

Extra-galactic OH megamaser emission is generally associated with compact ($<$
100 pc scale) starburst or AGN activity in the centres of IR luminous galaxies.
Observations of this maser emission provides a unique method of studying the
structure and kinematics of galactic nuclei at parsec resolution without dust
obscuration effects. Measurements of velocity gradients and line widths (e.g.
\cite{P01}, hereafter P01) already provide important constraints on stellar mass
densities and turbulent velocities in IR luminous galaxies.  Such observations
may also trace large scale obscuring tori in composite AGN/starburst sources
\citep{K03}. Potentially OH megamasers can also tell us about the size, density
and temperature of molecular clouds in the central ISM of starburst galaxies and
AGN. However, to accomplish these goals a better understanding of the OH
megamaser phenomenon is required.

Recent MERLIN and VLBI observations of OH megamasers have suggested that the
standard model, developed in the 1980s \citep{BAAN89}, might need modification
(see \cite{LSDL02} for a review). In this standard model OH maser emission is
generated by low gain ($|\tau|\la 2$) unsaturated amplification of background
continuum by a foreground OH amplifying medium. Although this medium is
comprised of discrete OH clouds it is implicitly assumed that there are many
such clouds and they individually have very low gains. These clouds therefore
form an effective gas in which statistical fluctuations in cloud number between
different lines of sight are unimportant and so the maser opacity varies slowly
across the source. Given these assumptions the amplifying medium is often
described as a diffuse screen.

Consistent with the standard model, early VLA and MERLIN observations showed
that OH maser and continuum emission overlapped. Since both continuum and maser
emission were assumed to be smooth on scales $\la 100$ pc VLBI observations were
not expected to reveal anything interesting. However, when VLBI observations
were finally made of Arp220, both compact continuum (\cite{D89}, \cite{S98b})
and compact OH maser emission were detected (\cite{LSDL94}, \cite{LSDL98}
hereafter L98). Remarkably, the bright maser spots in Arp220 were not coincident
with the continuum spots, and some maser spots displayed extreme
line-to-continuum ratios (i.e. $>800$ in L98). These observations were clearly
inconsistent with diffuse screen models. Slightly less extreme compact maser
emission was subsequently detected in other sources (\cite{T97}, hereafter T97,
\cite{D99}, hereafter D99, \cite{K03}, \cite{K04}). These same sources also
contain diffuse maser components which account for between 50\% and 90\% of the
total maser flux density. It has been suggested that the compact masers occur in
saturated, perhaps collisionally pumped regions, while the diffuse component
comes from an extended unsaturated, radiatively pumped screen fully consistent
with the standard model (L98, D99).

One of the clearest cases of an OH megamaser showing both compact and diffuse
maser emission is IIIZw35. Two groups of compact masers were detected in VLBI
observations (T98, D99), but only about half of the total OH maser emission seen
on MERLIN scales \citep{MC92} was recovered in the VLBI maps. To determine the
location of this missing component P01 conducted EVN+MERLIN observations. These
observations revealed perhaps the best example of a rotating OH maser ring yet
found. The previously known compact masers lie at the tangents of this ring.
Such a location is hard to understand if compact and diffuse masers are
generated by different physical mechanisms. Instead, a geometrical origin for
the compact masers is suggested. P01 proposed a mechanism based on a single
phase of OH masing small clouds (\about 1 pc). At the ring tangents multiple
overlaps between clouds in space and velocity are likely due to the increased
path length through the ring. These multiple cloud overlaps give rise to the
bright compact maser features. Elsewhere, at the front and back of the ring
where there are few such cloud overlaps, the emission consists of many weak
maser spots. These spots are too weak to be detected individually in high
resolution observations but in low resolution observations they are averaged
together and give rise to an apparently diffuse emission.

In this paper we investigate in greater depth the clumpy ring model proposed in
P01. Here we fully consider the spectral properties of the clouds and their
velocities are treated in a more realistic way. We also model the continuum
emission in a geometrically and physically consistent manner. Using numerical
simulations we demonstrate that most of the available OH maser observations of
IIIZw35 can be explained using an improved version of the P01 model. Most of the
input parameters for these simulations are constrained directly by the
observations. Our modelling also illuminates general properties of maser ring
geometries and clumpy maser media.

The detailed plan of this paper is as follows. Section \ref{se:observations}
summarizes the relevant observations of IIIZw35. In Section \ref{se:parameters}
we present the clumpy ring model and derive its parameters. Section
\ref{se:numerical} describes the numerical simulations and compares the results
of the simulations with observations. Section \ref{se:maser_properties} presents
a general discussion on how the observed maser characteristics can be explained
by a clumpy medium. In Section \ref{se:physics} the derived physical properties
of the clouds and the galactic nucleus are discussed. Finally in Section
\ref{se:conclusions} we draw conclusions and suggest future work.

\section{Source properties and observations\label{se:observations}}

IIIZw35 is a Luminous Infra-red Galaxy ($L_{\rm FIR} = \E{11.3} \Lo$) at a
distance of 110 Mpc (assuming H$_0 = 75$ \kms Mpc$^{-1}$, (\cite{CH90},
hereafter C90))  which gives a linear scale of 0.5 pc mas$^{-1}$. From optical
spectroscopy it has been classified as a borderline LINER/Seyfert 2 (C90). The
galaxy lies on the well known radio-FIR correlation and both the radio continuum
and IR emission are consistent with being powered by a starburst with a star
formation rate (SFR) of \about 19 \Mo yr$^{-1}$ (P01). Both the 1667 MHz and
weaker 1665 MHz megamaser emission were discovered by \cite{SS87}. The ratio of
peak opacities in the 1667:1665 MHz spectra is \about 9 (\cite{SS87} and 
\cite{MS87}), which is at the higher end of observations amongst OH megamaser
galaxies \citep{HW90}. From single dish CO observations, \cite{MS87} estimate an
overall molecular gas mass of 6.5\x\E{9} \Mo. The dynamical mass within a radius
of 22 pc is estimated by P01 to be \about 7\x\E{6} \Mo, implying that most of
the molecular gas is located in a more extended structure.

\begin{figure*}[ht!] \centering
\includegraphics[width=0.47\hsize]{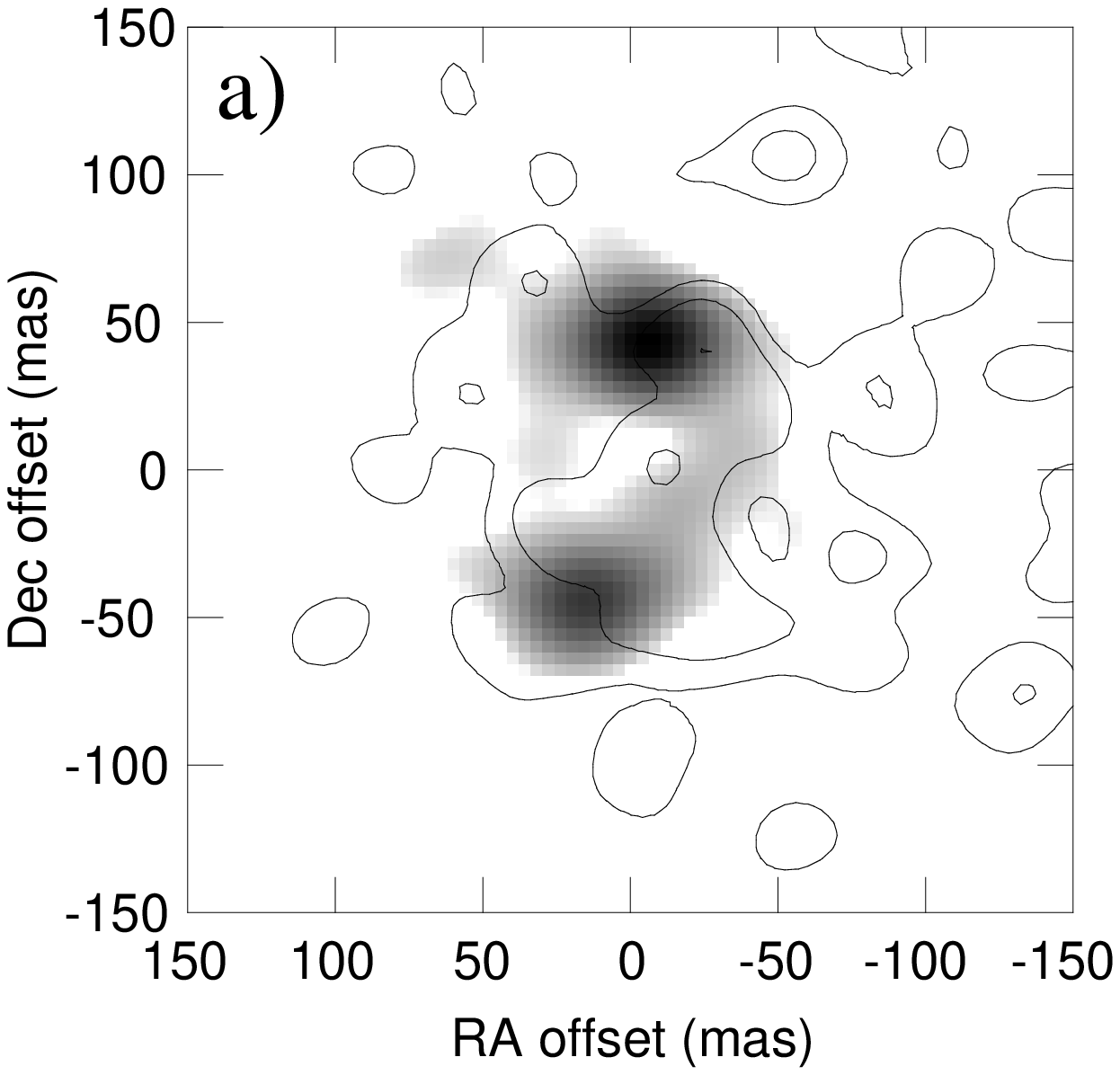} \hfil
\includegraphics[width=0.47\hsize]{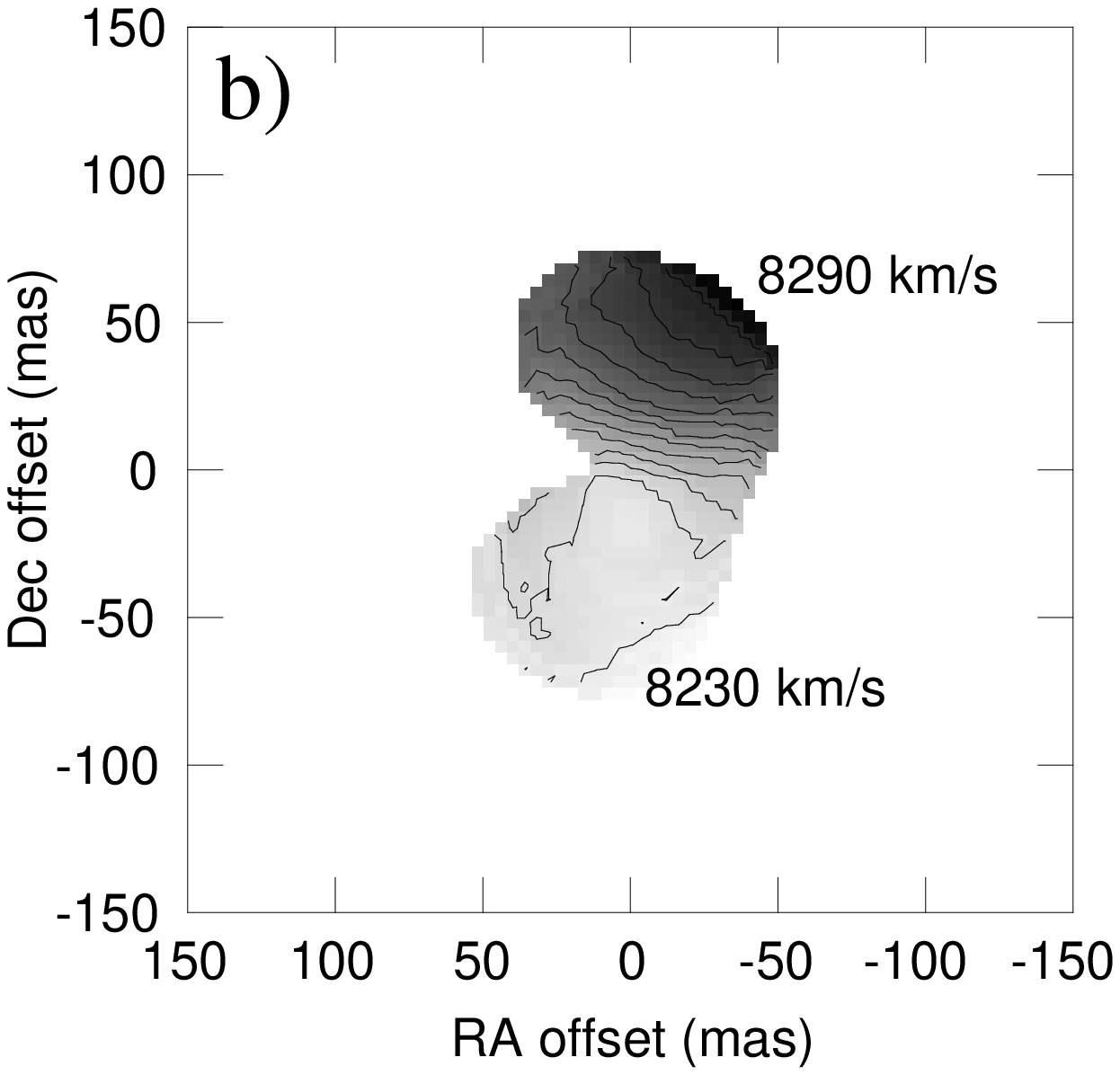} \caption{Images
reproduced from P01 showing observations of OH maser and continuum emission at 
EVN+MERLIN resolution {\bf a)} Greyscale shows the velocity integrated OH emission 
at a resolution of 34$\times$29 mas. The  contours show continuum emission at the 
same resolution. Contour levels are at -1,1,2, and 4 times the 3$\sigma$ noise
of 0.42mJy/beam. {\bf b)}  Corresponding OH maser velocity centroid field.
The greyscale is between 8224 (lightest grey) and 8300 \kms (black).  Contours
are from 8230 \kms\  increasing at 5\kms\ intervals  up to 8290 \kms .}
\label{fi:figure1} \end{figure*}

Intermediate resolution maps of the OH maser emission have been produced by P01
by combining EVN and MERLIN datasets and are reproduced in Figure 1. The
greyscale in Figure \ref{fi:figure1}a shows that, at this resolution, the maser
emission has a clear void near its centre and two bright regions to the North
and South, with an estimated Line-to-Continuum Ratios (LCR) of $47\pm13$ and
$73\pm8$ respectively. Two bridges of diffuse emission connecting the bright
regions can also be seen. Although the one to the West is brighter, it has the
smaller LCR; the western bridge has LCR = $9\pm2$, the eastern bridge has an LCR
= $14\pm4$. Figure \ref{fi:figure1}b shows the North-South velocity gradient
obtained from the maser moment map. This gradient and the maser brightness
distribution, are consistent with a rotating ring whose axis is inclined by $i$
\about\ 60\deg\ from the direction of the observer. The bright North and South
regions are interpreted as arising from the tangents of the ring, where path
lengths are longer, while the bridge emission originates at the front and back
sides of the ring (see Figure \ref{fi:figure2}). At VLBI resolution (T97, D99),
the diffuse bridges are not detected while the two bright regions break up into
complexes of bright maser spots. In one of the spots the inferred LCR is $>500$
(D99). In general the spots are unresolved, thus having sizes of less than $0.7$
pc (D99). Spot  spectral widths are measured to be 30-50 \kms (D99). Both, T97
and D99 find that the spot velocity centroids trace a roughly East-West gradient
within the northern region. Approximately half of the velocity integrated line
flux density seen in Figure \ref{fi:figure1}a within the northern and southern
regions comes from the compact spots, the rest is not detected at either VLBA
(T97) or global VLBI (D99) resolution.

In addition to the OH maser emission P01 also detected continuum emission in 
their EVN+MERLIN data (see contours in Figure \ref{fi:figure1}a). In contrast at
global VLBI resolution only upper limits to the continuum have been found (T97;
D99), implying that most of the continuum emission is smoothly distributed. The
continuum emission seen at EVN+MERLIN resolution, just like the OH emission, is
stronger on the western side of the source and is therefore asymmetrically
distributed about the major axis of the inferred OH maser ring.

\section{Clumpy ring model
\label{se:parameters}}

\subsection{Overall Geometry} \label{se:genmod}

\begin{figure*}[ht!] \centering
\includegraphics[width=0.54\hsize]{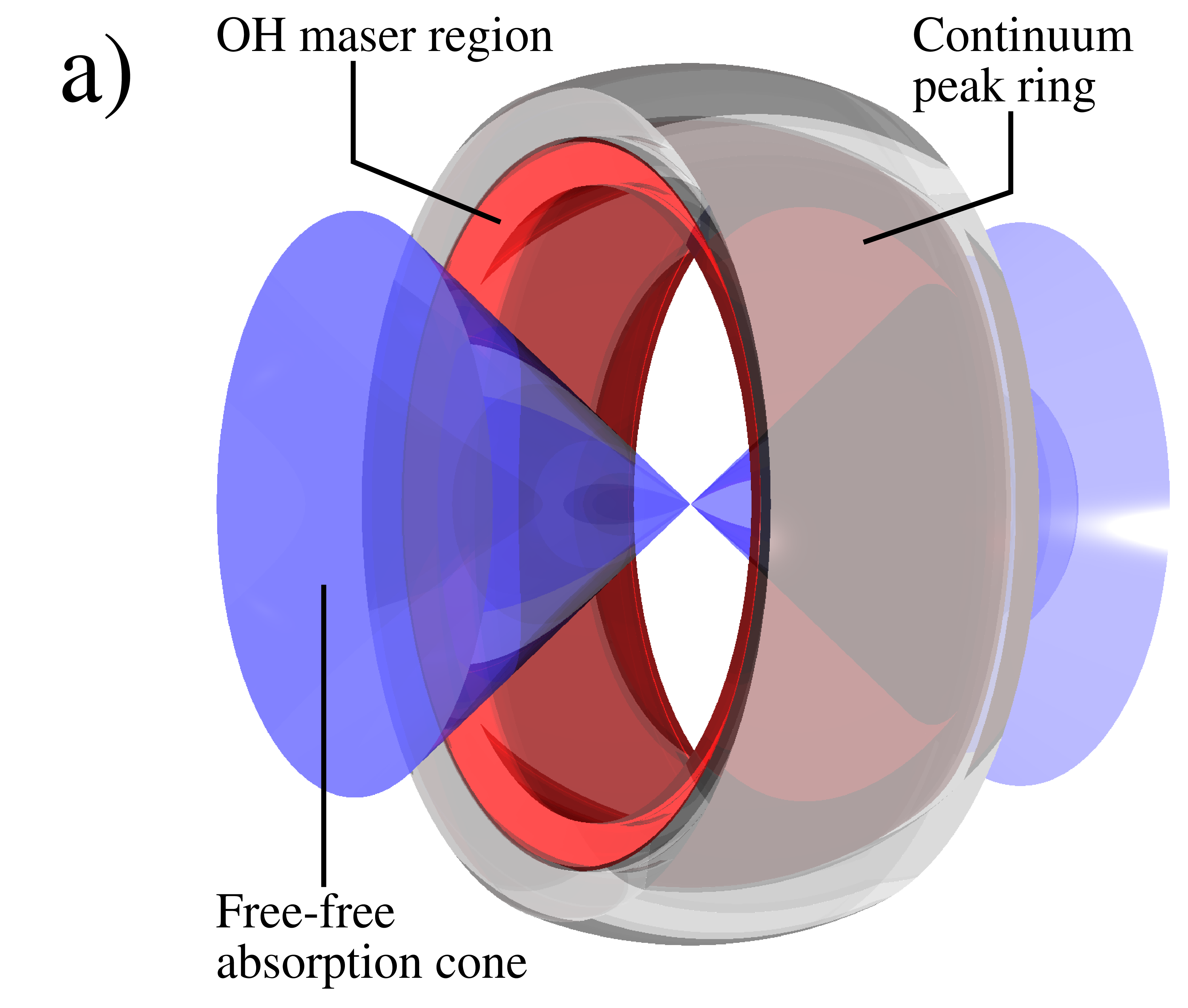} \hfil
\includegraphics[width=0.45\hsize]{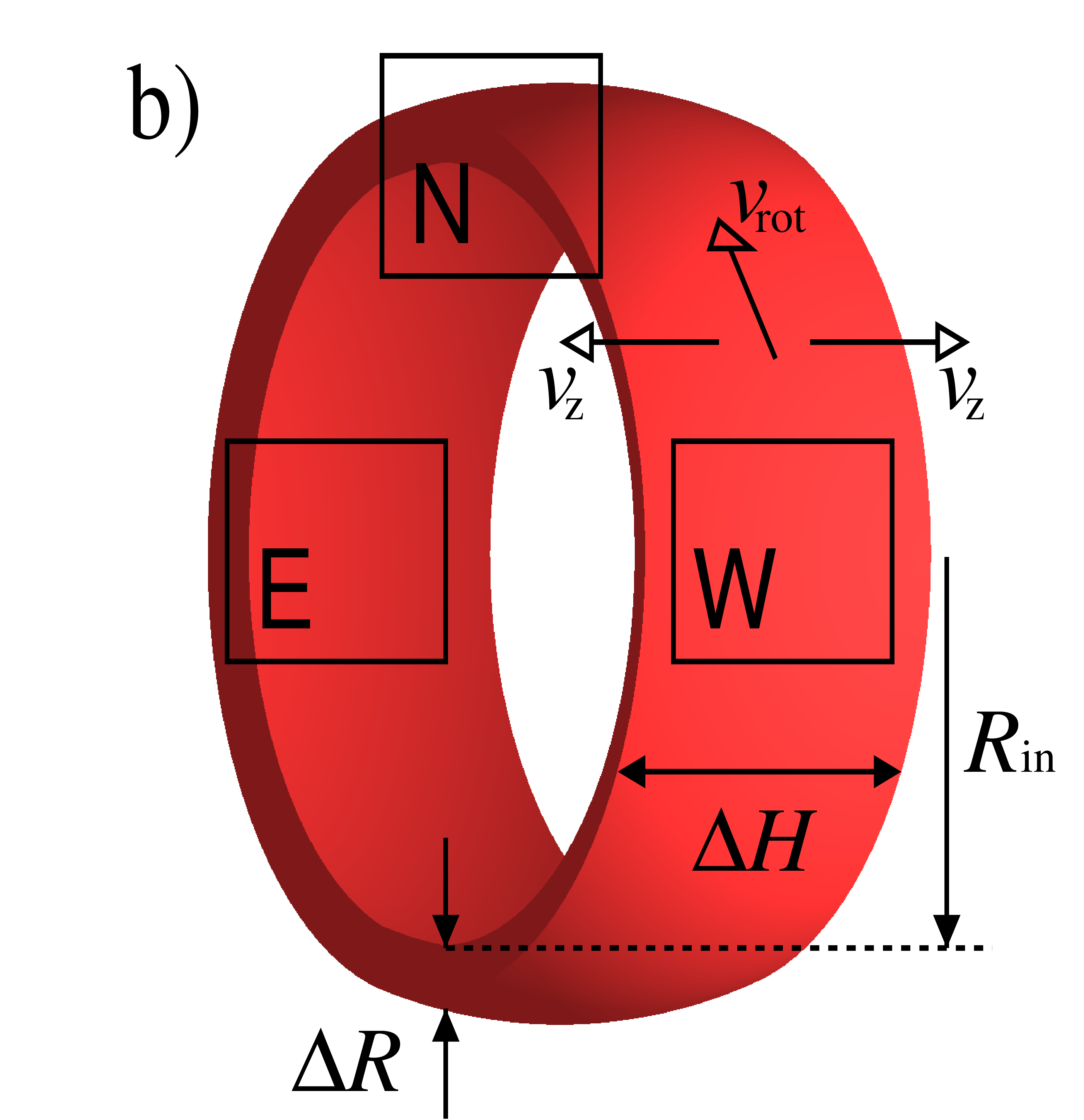} \hfil  \caption{{\bf a)}
Sketch of the proposed source geometry. The inner dark grey ring represents the
region where OH masing clouds are confined. The outer light grey ring depicts an
isosurface of the smoothly distributed continuum emissivity. Note that although
the radius of peak continuum emissivity lies outside the OH maser ring, some
continuum emission interpenetrates and even lies within the OH ring (not shown
in this figure). This geometry explains the large line to continuum ratio on the
eastern  side of the source because here the majority of the continuum is
background to the OH. In contrast on the western side only the smaller fraction
of the continuum which interpenetrates and lies inside the OH maser zone is
available  as a source of seed photons. In order to explain the relative
weakness of the absolute brightness of both line and continuum on the eastern
side of the source (see Figure \ref{fi:figure1}) the model also includes a
region of free-free absorption within a bicone which covers the eastern side of
the source. {\bf b)}: Detailed representation of the OH maser ring indicating
positions and dimensions referred to in the main text. Arrows indicate
components of cloud rotation around the ring ($v\sub{rot}$) and outflowing from
the ring midplane ($v\sub{z}$). See {\S\ref{se:velocity}} for more details. }
\label{fi:figure2} \end{figure*}

In this paper we fit the IIIZw35 observations by an inclined axisymmetric model
in which both OH clouds and continuum emission coexist within circumnuclear
rings (see Figure \ref{fi:figure2}). To reconcile such a symmetric geometry with
the observed East-West asymmetry in both line and continuum emission (see Figure
\ref{fi:figure1}a) our model also includes a bi-cone of free-free absorption
which covers the eastern side of the source. This obscuration defines the maser
ring orientation requiring the eastern side to be the most distant. Note that
the existence of a free-free absorbing component is supported by independent
evidence. C90 found that the integrated radio spectrum has a turnover at \about
1 GHz. Given the size of the radio emission this cannot realistically be due to
synchrotron self-absorption, because enormous departures from particle/field
equipartition would be needed. Physically the free-free absorbing cone could be
the base of an outflowing superwind such is often observed in energetic
starbursts \citep{HE03}.

As noted in \S\ref{se:observations} although the total maser emission is weaker
on the eastern side the LCR is about factor of 2 larger there than on the
western side. This difference is explained if more of the seed continuum
emission is background to the OH ring on the east side. This naturally occurs if
most of the bright continuum emission comes from a larger radius than the OH
masing gas (see Figure \ref{fi:figure2}a). This is corroborated by the fact that
the observed continuum radio source is larger than the megamaser source (see
Figure \ref{fi:figure1}a). As shown in \S\ref{se:results_and_comparisons} this 
geometry also predicts that the brightest masers do not occur exactly at the
maser ring tangents but slightly to the east of the tangent points, just as
found by the observations of T97 and D99.

The detailed properties of the different components of the model are estimated
in the following subsections and summarised in Table \ref{ta:table1}. These
parameters are used in the Monte-Carlo simulations described in 
{\S\ref{se:numerical}}.

\subsection{Free-Free absorption Bicones}\label{se:freecone}

From the observed ratio of the continuum emission brightness on the East and
West sides of the source, the line of sight opacity through the cone to the
eastern bridge is estimated to be $\about 0.7$ at the observing frequency. Since
all of the eastern bridge emission is weak compared to the western bridge the
opening angle of the bicone must be significant. We therefore assume a bicone 
in which the opacity  per unit length decreases with angle from the symmetry axis 
as a gaussian with  FWHM $60^{\circ}$.

\subsection{Ring dimensions}\label{se:dimensions}

The dimensions of the region containing the OH masing clouds are shown in detail
in the right panel of Figure \ref{fi:figure2}. This thin ring structure has
inner radius \Rin\ and an outer radius \Rout\ = \Rin\ + \DR. P01 find that \Rin\
\about\ 22 pc. From the North-South extent of the region covered by bright maser
spots in the VLBA and global VLBI maps (T97 and D99), \DR\ is \about\ 3 pc. From
its East-West extent near the ring tangents we infer a ring height of \DH\
\about\ 6pc. The aspect ratio of the projected ring shows that the ring axis is
inclined at angle $i$ \about\ 60\deg\ from the observer's direction (P01). We
find that an axisymmetric continuum emissivity profile which peaks at radius
$\Rc =37$ pc best fits the continuum observations (see Figure \ref{fi:figure1}a
and \S\ref{se:model_computation}). We also find that the continuum emissivity
function must be relatively wide in radius so that not all of the continuum lies
outside the OH maser region.

\def\n#1{\hbox{$^{\rm #1}$}}
\begin{table}
\caption{Model Parameters  \label{tab:tab1}} \centerline{}
\begin{tabular}{ll}
\hline \hline

\noalign{\medskip}
{\em Maser and continuum ring properties}\/: &                    \\

\sp Maser ring inner radius \Rin\n{a}          	&22 pc             \\
\sp Maser ring radial thickness \DR\n{a}       	&3 pc             \\
\sp Maser ring height \DH\n{a}                 	&6 pc              \\
\sp Maser ring Rotation velocity $v\sub{rot}$\n{b}	&57 \kms           \\
\sp Continuum peak emissivity radius \Rc\n{b}  	&37 pc             \\
\sp Inclination angle $i$\n{b}      	&60\deg            \\

\hline \noalign{\medskip}
{\em Free-free absorption bicone properties}\/:    &                \\
\sp Opacity $\tau\sub{ff}$    		& 0.7      \\
\sp Opening angle (FWHM)                & $60^{\circ}$      \\

\hline \noalign{\medskip}
{\em Cloud properties}\/:    &                \\

\sp Cloud size $a$\n{c}                   		& $<$0.7 pc       \\
\sp Internal velocity width \dv\n{c}\n{,d}      	& 20 \kms    \\
\sp Random velocity dispersion \Dv\n{c}\n{,d}		& 60 \kms    \\
\sp Outflow velocity $v_{z}$  				& 60 \kms    \\
\sp Outflow velocity dispersion $\Delta v_{z}$\n{d}	& 30 \kms    \\
\sp Number of clouds along radius \No   & 1.8          \\
\sp Cloud volume filling factor $f$		& $<0.08$	\\
\sp Cloud maser optical depth $\tau\sub{c}$   & 1.5          \\
\hline \hline
\\
Refs: \n{a}T97, \n{b}P01,  \n{c}D99  &                 \\
\n{d} The dispersions represent the FWHM of a
\\gaussian probability distribution.
\end{tabular}\label{ta:table1}
\end{table}


\subsection{Need for a single-phase clumpy medium} \label{se:need4clumps} 

Having constrained the geometry of the maser emitting zone we now argue that it
must consist of distinct clouds. Maser amplification of background continuum
gives LCR = $e^\tau$, where $\tau$ is the maser optical depth. When the
amplification is due to a smooth distribution of uniform unsaturated maser gas,
$\tau$ is proportional to the path-length through the region. The ring geometry
derived in the previous sections implies a factor of 8 difference between the
maximum path-lengths in regions N and E (see Figure \ref{fi:figure2}) such a
ratio in path length implies LCR$\sub{N}$ = (LCR$\sub{E}$)$^8$. In contrast,
the ratios observed at EVN+MERLIN resolution are LCR$\sub{E}$ \about\ 12 and
LCR$\sub{N} <75$, a conflict of at least four orders of magnitude with this
relation. The problem becomes even worse if saturation affects the bright
features because the optical depth of a saturated maser is always larger than
that of an unsaturated maser of the same length.

In contrast with a smooth maser screen, a collection of independent clouds
involves a statistical effect, predicting a variation in LCR with position
around the ring which can match the EVN+MERLIN observations. As described in the
appendix, a remarkable property of clumpy media is that the effective opacity is
not linearly proportional to the path length, instead, it falls below the linear
prediction. This result gives a compelling argument for the OH  masers occurring
in clouds. This argument is separate and additional to the one proposed in P01
which required clouds to explain the brightest spots by multiple cloud overlaps.

\subsection{Cloud Properties} \label{se:cloudprops} 

Based on the high resolution observations of the brightest maser spots in D99
maser spot diameters are estimated to be $\lesssim 0.7$ pc. In general the spot
size associated to a high gain maser is smaller than the size of its originating
maser cloud (by a factor of \about$\sqrt{\tau}$). However since the maser
opacity of the clouds employed in our model is about one (see below) the size of
the maser produced by a single cloud is about the same size of the cloud. In
spots caused by multiple cloud overlaps the apparent reduction in size caused by
increased opacity is counteracted by the widening due to spatial misalignments.
The cloud size is therefore estimated to be the same as that set by spot size,
$<0.7$pc. A lower limit for the cloud size is set from physical arguments (see
\S\ref{se:physics_of_clouds}). For our Monte-Carlo simulations
(\S\ref{se:numerical}) we adopt the observed upper limit as the actual cloud
size but argue that our results are only weakly dependent on the exact size
chosen.

From the observed velocity widths of the bright maser spots (D99) cloud internal
velocity dispersions of order 20 \kms\ are implied. Again a similar argument
applies in velocity as in space to explain why we expect observed and true cloud
velocity widths to be similar. Values for the cloud number density and opacity
were found from the Monte-Carlo simulations. As described in
\S\ref{se:model_details}, the initial starting point for this search was guided
by the observed LCR distribution around the ring (see \S\ref{se:need4clumps})
and the theory developed in the appendix. The best fit was obtained assuming a
cloud opacity of $\tau_{c}=1.5$ and a volume number density such that the number
of clouds along a radius in the disk plane at any velocity was $N_{o}=1.8$. For a cloud
size of $a<0.7$pc this implies, averaged over the OH maser ring, a 
 cloud volume filling factor $f<0.08$. As we argue in 
\S\ref{se:model_details} our fits to the data are expected to remain similar as 
a function of $a$ as long as  the cloud number density is adjusted to maintain 
a constant  $N_{o}$. In this case the cloud volume filling factor will
scale with cloud size as  $f=0.08(a/0.7\rm{pc})$.

\subsection{Velocity field}
\label{se:velocity}

Observations show (see Figure\ref{fi:figure1}b) a large scale North-South
gradient in maser velocity centroid consistent with ring rotation. Note however
that the velocity dispersions within the bright northern and southern tangent
regions are comparable to the rotation velocity so that the integrated spectra
from the two regions overlap (see P01 and T97). At higher resolution global VLBI
observations show within the northern tangent region a clear velocity gradient
which is almost East-West. This small scale gradient is almost parallel to the 
projected ring axis and almost at right angles to the gradient found on large
scales.

The observed large scale gradient can readily be fitted if the clouds are
assumed to have an orbital velocity component (see Figure \ref{fi:figure2}b) of 
$v\sub{rot}=57$ \kms\ (P01). Somewhat harder to explain is the gradient seen
within the northern tangent region. Part of this gradient can be explained by
the ring rotation. The brightest masers are not found exactly at the OH maser
ring tangents but slightly to the East where there is more background continuum
(see Figure \ref{fi:figure2}a) and the orbital velocity gives a small velocity
gradient along the ring. This mechanism cannot however explain the magnitude of
the observed gradient. Instead we propose that clouds, in addition to their
rotation velocity, have a comparable outflow velocity ($v\sub{z}$, see Figure
\ref{fi:figure2}b) which is parallel to the ring axis and directed away from the
disk midplane. To see how this gives the observed gradient consider a cloud near
a tangent region in Figure \ref{fi:figure2}b. If such a cloud lies above the
ring midplane (near side), the cloud will be located to the East of the tangent
point and its projected outflow velocity will be blueshifted. On the other hand
if it lies below the ring midplane (far side) the cloud will be located to the
West of the tangent point and its projected outflow velocity will be redshifted.
This mechanism can produce an apparent East-West velocity difference of nearly
twice the projected outflow velocity within a small distance comparable to the
maser ring height.

It is interesting that at the northern tangent the cloud outflow mechanism gives
a velocity gradient in the same direction as that caused by the rotation
mechanism, thus reinforcing the gradient. In contrast in the South the gradients
from the two mechanisms have opposite directions and will partly cancel. This
may explain why an East-West gradient is only seen in the North. Alternatively
the difference may mainly be due to the statistical nature of our clumpy maser
model. We have found using Monte-Carlo experiments (see \S\ref{se:numerical})
and assuming an outflow velocity of $v\sub{z}=60$ \kms\ and dispersion $\Delta
v\sub{z}=30$ \kms\ that small scale gradients of the required amplitude often
arise on the tangent regions, with the northern region being favoured. 

While the outflow velocity component contributes significantly to the high
velocity dispersion seen at EVN+MERLIN resolution we also find from our
Monte-Carlo modeling that an additional cloud random velocity component of
dispersion $\Delta v=60$ \kms\ is required. In particular this random velocity
component is needed to fit the observed very wide velocity dispersion of the
northern and southern tangent regions at EVN+MERLIN resolution.

\section{Modelling and results}\label{se:numerical}

In this section we calculate in detail the emission from the clumpy maser ring
using Monte Carlo simulations. In \S\ref{se:model_details} we present the
details of the model. In \S\ref{se:model_computation} we explain how we compute 
our synthetic  OH maser cubes  and continuum images. Finally in
\S\ref{se:results_and_comparisons} we compare our results with observations.

\subsection{Model Details}\label{se:model_details}

Our numerical model contained three components (see \S\ref{se:genmod}): a high
brightness temperature radio continuum emission to provide seed photons for
maser amplification, a free-free absorbing cone and a randomly distributed
population of OH masing clouds. The continuum component was modelled as a
radially smooth axisymmetric emissivity function. This function was chosen to
have peak emissivity at radius $R\sub{c}$ (see Table \ref{ta:table1}) and its
shape was chosen so that it approximately matched the P01 continuum image (see
Figure \ref{fi:figure1}a). The bi-cone of free-free absorption (see Figure
\ref{fi:figure2}a) was assumed to have a density obeying a Gaussian distribution
of FWHM=60$^{\circ}$ around the ring axis producing a total opacity of $\tau =
0.7$ toward the eastern bridge region (see \S\ref{se:freecone}).

\begin{figure*}
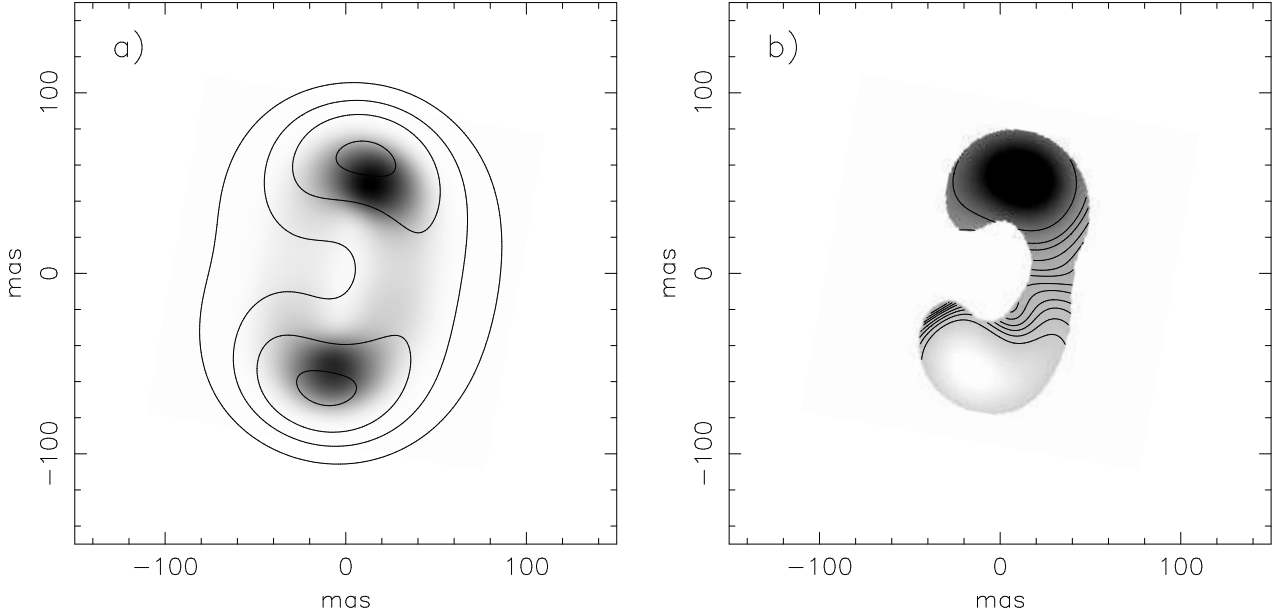
\centering
\includegraphics[width=0.45\hsize]{figures/2971fi3a.ps}\hfil
\includegraphics[width=0.45\hsize]{figures/2971fi3b.ps} \caption{{\bf a)} Model
continuum emission superimposed on the velocity integrated OH maser emission.
The contours, greyscale and the resolution are equivalent to those in Figure
\ref{fi:figure1}a. {\bf b)} Corresponding modelled OH maser velocity centroid
field. Greyscale is between -60 and 60 \kms around the systemic velocity.
Contours are from -30 \kms\ and increasing by 5 \kms\ up to 30\kms\ . Compare
with Figure \ref{fi:figure1}b. } \label{fi:continuum} \end{figure*}

The maser clouds were assumed to be identical and spherically symmetric with a
number density of maser molecules described by a three dimensional Gaussian
distribution with a FWHM of 0.7pc (see \S\ref{se:cloudprops}). For computational
reasons this density profile was cutoff beyond a radius of 1pc. The
corresponding cloud internal velocity profile was assumed to be a Gaussian with
FWHM \dv\ = 20 \kms.

As discussed in \S\ref{se:physics_of_clouds} the clouds could be physically
smaller than the above FWHM of 0.7pc and still give the required maser opacity.
Fortunately we find that when the clouds are small enough to be unresolved by
the highest resolution interferometer beam the model results are only very
weakly dependant on the actual cloud size used. As described in the appendix the
critical parameters for determining structure are the cloud opacity $\tau_{c}$
and mean number of clouds per line of sight $\Nbar$. If a smaller cloud size was
used and consequently the volume density of clouds increased to maintain the
same $\Nbar$, then the results to first order would be the same. The only
explicit dependence on cloud size is via the inverse Poisson function (see
appendix) which is an extremely weak function of the number of independent lines
of sight ($M$); a quantity which in turn is inversely proportional to the cloud
area. The effect of using clouds much smaller than used in our simulations would
be to make the peak line to continuum and 1667:1665MHz line ratio somewhat
larger, but as we shall see (see \S\ref{se:results_and_comparisons}) these ratio
limits are well fitted even when using the largest possible cloud size.

As described in \S\ref{se:velocity} the cloud bulk motions comprise orbital,
outflow and random components. To implement these motions in our simulations  we
first found for each cloud six orbital parameters for a random orbit around the
gravitational potential induced by the 7\x\E{6}\Mo\ enclosed mass. The maximum
allowed orbit inclination was set to the ratio \DH/2\Rin.  Clouds were also
constrained to lie within an annulus with inner and outer radii \Rin\ and
\Rout=\Rin + $\Delta R$ (see Table \ref{ta:table1}). The outflow from the ring
plane (see \S\ref{se:velocity}) was simulated by adding a velocity component
parallel to the ring axis to each cloud. After experimenting it was found that
assuming a mean $v_z=60\kms$ with dispersion $\Delta v_z=30\kms$ directed away
from the ring plane, plus a random 3D velocity of 60\kms gave acceptable
results. This velocity field produced realisations which showed velocity
gradients within the northern tangent region and could also match the observed
velocity dispersion.

As described in \S\ref{se:cloudprops}, the volume number density of clouds
$N\sub{c}$ and their opacity $\tau\sub{c}$ were left as adjustable parameters
within the Monte-Carlo simulation. To guide the search for these parameters we
utilised the formalism developed in the appendix. A system of two equations was
set up using the expressions for the beam-averaged gain over the northern and
eastern regions. The system was solved in the least-squares sense subject to a
lower-bound constraint established by the observed peak gain at global VLBI
resolution. The optimum solution found had $\tau\sub{c}=1.5$ and a cloud number
density in the midplane of $N\sub{c}=0.2$ pc$^{-3}$. For this density the number
of clouds intersected at any velocity by a radial equatorial ray is \No=1.8. The
required number density was achieved in our simulations by using 1200 clouds.

\begin{figure*}[ht!]
\includegraphics[angle=-90,width=1\hsize]{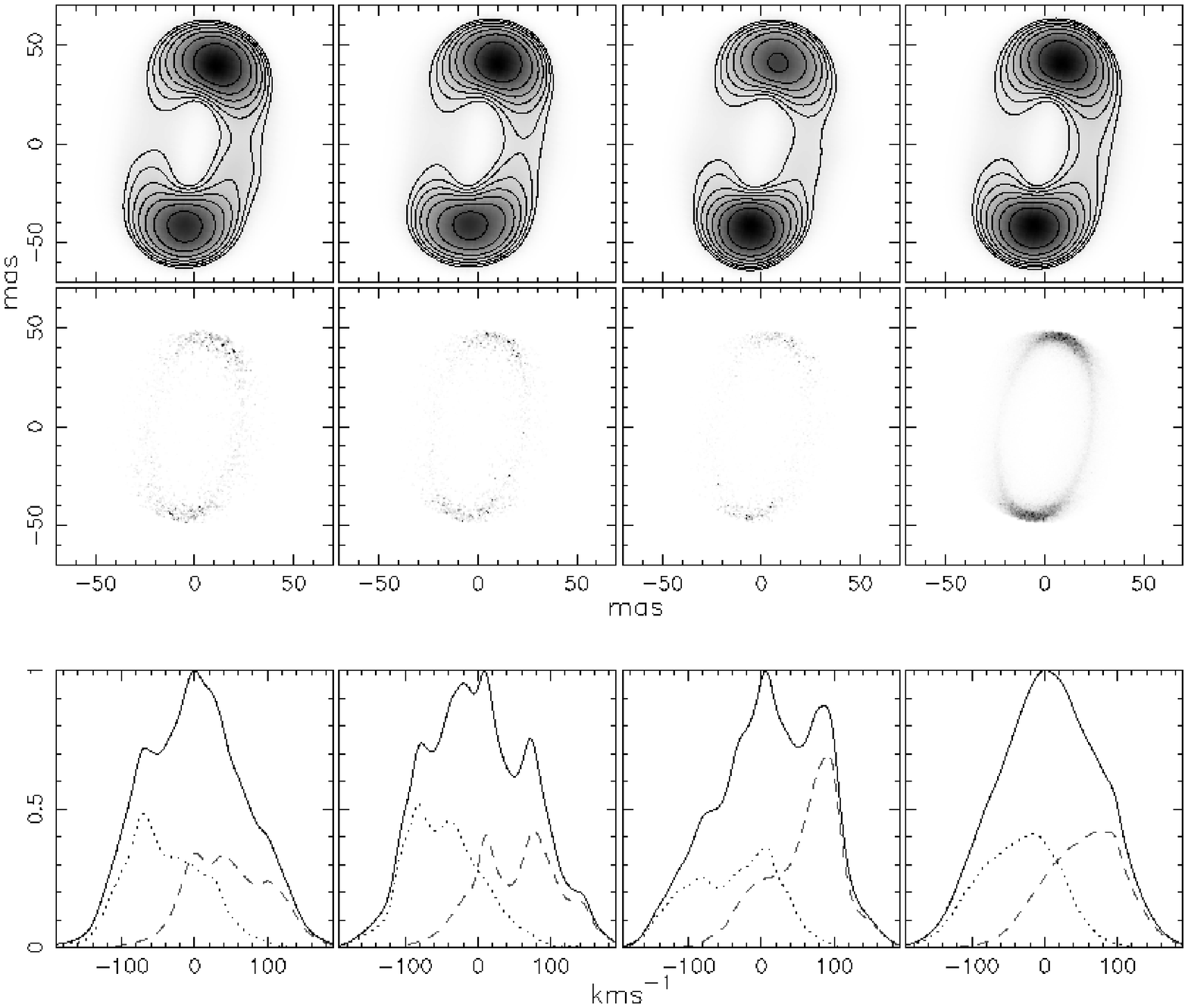}
\caption{The first three columns show sample realisations of the Monte-Carlo 
simulation. The rightmost column shows the average over 100 realisations.
Intensity and flux are in arbitrary units, normalized to their peaks.  {\bf
Images:} Top Row: Greyscale and contours represent the velocity integrated OH
maser emission convolved with a simulated EVN+MERLIN beam (FWHM $\approx$ 35
mas). Contours are logarithmically spaced from 2$^{-1/2}$ to 2$^{-7/2}$ of the
peak. Middle Row: Integrated emission at global VLBI resolution (FWHM = 0.92
mas). {\bf Spectra:} The solid line is the spectrum integrated over the entire
source. Dashed and dotted lines represent spectra taken at the southern and
northern tangent regions respectively.} \label{fi:realizations1} \end{figure*}

\begin{table}
\caption{Observed and modeled quantities}
\begin{tabular}{lrr}
\hline \hline \noalign{\medskip}

                             &Observed        &Model\\

\hline\noalign{\medskip}
{\em Single dish}\/:         &                &\\
\sp Velocity range\n{a} (\kms)&270 $\pm$20     &\about 280\\
\hline\noalign{\medskip}
{\em EVN+MERLIN}\/:          &                &\\
\sp $\tau\sub{North}$\n{b}   &3.85$\pm$0.34   &3.58\\
\sp $\tau\sub{South}$        &4.30$\pm$0.12   &3.80\\
\sp $\tau\sub{East}$         &2.66$\pm$0.33   &1.20\\
\sp $\tau\sub{West}$         &2.26$\pm$0.21   &0.69\\
\hline\noalign{\medskip}
{\em Global VLBI}\/:         &                &\\
\sp $\tau\sub{South}$        &$>$6.21         &7.28\\
\sp FWHM of peak spot (\kms) &\about30        &30\\
\hline \hline

\multicolumn{3}{l}{\n{a}Velocity width of the 1667 feature at 10\% of the peak}\\
 flux density.\\

\multicolumn{3}{l}{\n{b}The effective opacities are $\tau\sub{North} = {\rm
ln}({\rm LCR}\sub{North})$, etc.}

\end{tabular}
\label{ta:table2}
\end{table}

\subsection{Model Computation}\label{se:model_computation}

Maser emission was calculated by ray tracing for each line of sight (LOS) and
velocity taking into account the amplification of continuum seed photons by the
interspersed maser clouds. Synthetic spectral line data cubes and continuum
images were produced and convolved with appropriate beams for comparison with
observations. The simulations were run in a dual Pentium III 1.5GHz computer
with 2GB of RAM running Linux. A realisation using a synthetic cube of size
201$^{3}$ with 1200 clouds took approximately 20 min. An important aspect of our
implementation is that for each realisation, the cloud information is stored as
a list of real numbers (and not gridded), so it is relatively easy and efficient
to recompute the expected emission at different resolutions by specifying
different grid sizes.

\subsection{Results and comparison with
observations}\label{se:results_and_comparisons}

A set of 100 simulations were run using the parameters given in Table
\ref{ta:table1}. The maser and continuum emission from a typical realisation are
shown in Figure \ref{fi:continuum} in a way that can be directly compared to the
observations in Figure \ref{fi:figure1}. Relevant numerical results for this
realisation compared to observations are given in Table \ref{ta:table2}. The
model continuum emission shown by the contours in Figure \ref{fi:continuum}a
matches well the high brightness regions of the observed continuum (compare to
contours in Figure \ref{fi:figure1}a). It is clear that our continuum model does
not include the very extended emission present in the observations. However we
consider that this model is sufficiently accurate for our main purpose of
calculating synthetic OH maser emission cubes. 

The greyscale in Figure \ref{fi:continuum}a shows the integrated OH maser
emission from our model and it can be seen to be very similar to the
observations (compare to Figure \ref{fi:figure1}a). Figure \ref{fi:continuum}b
shows the model velocity field obtained from the moment map of the OH emission.
Again, its overall structure agrees quite well with observations (compare to
Figure \ref{fi:figure1}b). There is a North-South asymmetry in the model
velocity field which in this realisation is partly caused by the presence of a
single bright maser feature in the South. The outflow velocity given to the
clouds also contributes to break the North-South symmetry (see
\S\ref{se:velocity}). The observations (Figure \ref{fi:figure1}b) show a
similar but larger asymmetry between North and South. This might be explained by
the fact that in the observations the south contains an even more dominant
single spot than is the case in the model. Additionally note that the model
moment map is produced using all the maser emission from the high resolution
cube whereas in the observations there may be a resolved-out component not
included in the average.

In order to show the range of structures produced in the Monte-Carlo simulations
Figure \ref{fi:realizations1} displays three selected realisations of the model
and the average over the entire set. The top two rows show velocity-integrated
images at MERLIN+EVN and global VLBI resolutions, the bottom row shows the
corresponding integrated spectra. As expected, the simulations show bright
emission at the tangent regions and diffuse emission in between. The intensity
is lower in the eastern bridge due to the effect of the free-free absorbing
cones, yet the LCR is higher there, reflecting the relative locations of the
maser clouds and radio continuum. From Table \ref{ta:table2} we see that the
values obtained for the EVN+MERLIN optical depths in the bridges are somewhat
less than observed. This may be due to the fact that the model uses only one
type of spherical cloud representing a whole population that certainly has
various sizes and optical depths. The bridge regions have small effective
optical depths and can be expected to be more strongly affected by fluctuations,
which are amplified exponentially, around the mean cloud optical depth. At
global VLBI resolution (Fig \ref{fi:realizations1} middle row) the bright
tangent emission breaks up into numerous compact maser spots due to multiple
cloud overlaps. With the greyscale level chosen there are more such spots on the
western side of the projected ring major axis, this is caused by the effect of
the free-free absorbing bicone.

The single dish model spectra (Fig \ref{fi:realizations1} bottom row) agree in
shape and velocity width with the \cite{SS87} observations. The spectra averaged
over the northern and southern tangent regions (dashed and dotted lines) are
also similar to those found by T97, including the velocity overlap of the two
regions; this overlap is due to the large cloud velocity dispersion (comparable
to the orbital velocity) included in the calculations. The emission from the
northern and southern tangent points comprises about half of the total emission,
as found in T97. The remaining missing flux comes from the eastern and western
bridges of smooth emission mostly centered on the systemic velocity, as found in
P01. The last column  of Fig \ref{fi:realizations1}, representing the average
over all the 100 realisations, shows that the model is stable. That is to say 
that the main structures produced, which match observations, are typical of most
realisations and are not unusual cases.

\begin{figure}\centering
\includegraphics[width=0.8\columnwidth]{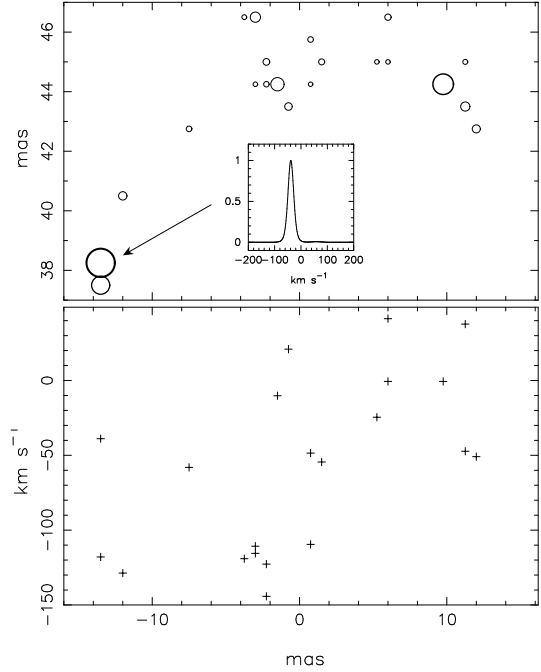} \caption{{\bf Top:}
Detail of the northern region of the best matching model realisation at global
VLBI resolution. The diameter of the circles plotted are proportional to the
maser spot velocity integrated emission. For comparison with the results of D99,
only spots brighter than 5\% of the peak are shown. The inset is the spectrum of
the brightest feature {\bf Bottom:} Position-Velocity diagram of the brightest
maser spots. The data suggests the presence of a linear gradient \about
30$\pm$12 \kms\,pc$^{-1}$ for spots on the eastern side of the plot.}
\label{fi:gradient} \end{figure}

Figure \ref{fi:gradient} shows in more detail the northern tangent region for
the best matching realisation. The top panel shows good overall agreement in the
number, distribution and sizes of spots as compared with the observations of D99
and T97. Amongst these very brightest masers, just as in the  observations,
there are more maser spots on the eastern side of the tangent point than the
western side. This is explained by the fact that on the eastern side there is
more background continuum (see Figure \ref{fi:figure2}a). Note that the middle
row of Figure \ref{fi:realizations1} shows an opposite behavior on larger
scales caused by the presence of the free-free absorption bicone. In a given
realisation (see  inset in Figure\ref{fi:gradient}) a typical spot spectrum has
FWHM of 30 \kms and FWZI \about 100 \kms, which is in agreement with the
observations presented in D99 and T97. The brightest spot shown in Figure
\ref{fi:gradient} is caused  by the partial overlap of 5 individual clouds. This
spot has an LCR of \about 1200 and a 1667:1665 MHz line ratio $\rho$ = 28. Both
values are consistent with the limits LCR $>500$  and $\rho > 20$ found by D99
in their brightest spot (see Table \ref{ta:table2}).

The position-velocity diagram of the compact maser spots shown in the bottom
panel of Figure \ref{fi:gradient} shows a linear gradient of \about 30$\pm$12
\kms\,pc$^{-1}$ for the eastern group of spots. This is similar to the gradient
of 32 \kms\,pc$^{-1}$ found by D99 in their central group of maser spots
(designated N2 in D99). In this model realisation no similar gradient was seen
in the southern tangent region; just  as is the case in the observations. This
illustrates the stochastic nature of the mechanism for producing velocity
gradients as described in \S\ref{se:velocity}. While our model can reproduce the
overall gradient seen in the northern tangent region the dispersion we find
about this gradient is significantly larger than that observed by D99. There are
several possible reasons for this. First it may be that better fits can be found
using models with larger mean outflow velocities (i.e. larger  $v_{z}$ in Table
\ref{ta:table1}) and smaller random components (i.e. smaller $\Delta v$ and
$\Delta v_{z}$ in Table \ref{ta:table1}). Secondly the outflow velocity field we
used was very crude, simply a change of sign in velocity for clouds on either
side of the midplane. If clouds are instead gradually accelerated with
increasing z-distance then tighter gradients might be produced. An interesting
possibility is that such apparent accelerations could occur if clouds have
limited lifetimes before their dissipation (see \S\ref{se:physics_of_clouds}).
In this case clouds emitted from the  midplane with higher $v\sub{z}$ would be
observed with systematically larger z-distances. Finally our assumption of
spherical clouds is unlikely to be true in reality. More systematic gradients
might be produced if maser features are instead due to overlapping filaments.

\section{Discussion of maser properties}\label{se:maser_properties}

\subsection{High brightness temperature and maser saturation}

Previous studies have argued that compact megamaser features imply saturated
amplification because of the absence of detectable seed continuum and the
resulting large LCRs ($>500$ in IIIZw35 and $>800$ in Arp 220; D99, L98). The
high brightness temperatures of the compact spots, $T\sub{b}>2\x\E{9}$ K in
IIIZw35 have also been taken as evidence for saturation (D99). However, even an
LCR of 800 implies maser amplification with $\tau$ of only 6.7, considerably
below the typical saturation requirements of astronomical masers, $\tau$ \about
12--15 \citep{MOSHE92}. In our model, the brightest features arise from the
alignment of $n$ = 5 maser clouds each with $\tau_{c} = 1.5$ giving  a total
optical depth of 7.5, well below that required for saturation. The assumption of
unsaturated amplification made in the modelling is therefore self consistent
even for the strongest features.

It is also important to remember that saturation depends on the angle-averaged
intensity  $J = I\x(\Omega/4\pi)$, so even larger brightness temperatures and
LCRs are possible without saturation whenever the beaming angles $\Omega$ are
small. Such small beaming angles are natural in the case of a collection of
small  clouds as we have in our model. If there are $n$ spherical masers, each of
optical depth $\tau_c$ and diameter $a$ aligned along a length $L$ then

\eq{\frac{\Omega}{4\pi} = {1\over16n\tau_c}\left({a\over L}\right)^2 }

With $a$ = 0.7 pc, $\tau_c=1.5$ and $L$ \about 13 pc at the ring tangents, then
the brightest maser spots, for which $n=5$, have $\Omega/4\pi$ of only \about\
2.4\x\E{-5}. Hence maser spot brightness temperatures many orders of magnitude
larger than the present limits would still be consistent with unsaturated
emission. It is important to note that the tight beaming produced by lines of
overlapping maser clouds is highly anisotropic. The large LCRs and brightness
temperatures achieved in our model would not be observed from locations along
the ring axis.

\subsection{Maser line ratios}

Compact masers features observed in OH megamasers, display large values of $\rho
= I_{1667}$:$I_{1665}$. For instance, in Arp 220, L98 find from their VLBI
observations $\rho > 100$ in the brightest maser spots, in contrast with single
dish spectra where $\rho$ is only 4.2. In IIIZw35, the 1665 MHz line was
detected only toward one bright feature at VLBI resolution (D99), yielding $\rho
> 20$. In all other features, this maser emission is below detection.

The large observed values of $\rho$ were a major impetus for the proposition
that the compact emission is different in nature from the diffuse maser
component. However, the observed difference in $\rho$ between these two apparent
phases is a natural property of clumpy unsaturated masers. If $\tau$ is the 1667
MHz optical depth of a single cloud and $R = \tau_{1667}$:$\tau_{1665}$, then
$n$ overlapping clouds will produce $\rho = \exp(n\tau\cdot[1-1/R])$. From an
extensive compilation of OH megamaser sources, \cite{HW90} find that $R = 1.9
\pm 0.3$ for the sample average, similar to the ratio of 1.8 if the two
transitions have the same excitation temperature. For the five-cloud overlap
responsible for the brightest feature in our simulation, $\rho > 20$ for $R >
1.7$, consistent with the observations. The line ratio of the velocity
integrated emission from our simulated model single dish spectra is \about 4
which is comparable to the observed value from Arecibo spectra of 6.1 
\citep{MS87}.

\subsection{Broad maser lines}

In order to fit the observed velocity width of the compact maser spots, our
model requires clouds with an internal velocity dispersion of 20 \kms. Although
fairly large, this is significantly less than if each compact maser were
produced in a single unsaturated cloud, in which case line narrowing due to the
high amplification would require cloud velocity widths of \about\ 60 \kms. In
our model, line narrowing due to unsaturated maser amplification is largely
canceled out by velocity broadening from combining many clouds with different
velocity centres.

In bright galactic masers, the brightest spectral features are often quite
narrow ($<1$ \kms). This might be expected from a population of individually
emitting maser clouds. In that case, clouds with small velocity width would give
the strongest masers because the inverted column density is spread over a narrow
velocity range so these clouds have the largest peak gains. In contrast, the
linewidths of bright megamaser spots are typically several tens of \kms, and up
to 150 \kms\ \citep{LSDL02}.  Such large widths cannot arise from a population
of narrow-width clouds because they would require unrealistic multi-cloud
superpositions. These large widths clearly indicate spectral blending of
individual clouds with internal velocity widths $\gg$ 1 \kms. Such blending
arises naturally from cloud overlaps in a population with a distribution of
internal cloud properties. Multiple cloud maser emission requires overlaps in
both space and velocity, therefore narrow-velocity clouds, which rarely overlap,
are selected against. The overlap of a small number of clouds with large
internal velocity dispersion, and thus small peak gains, naturally leads to
broad spectral lines of moderate gain.

\subsection{Application of the model to other sources}\label{se:othersources}

Does the model proposed here apply to other sources showing both compact and
diffuse maser components? The best studied OH megamaser source is Arp220,
observed at global VLBI resolution (L98) and with EVN and MERLIN separately
\citep{RO02}; though there is no published map using the combined data from both
arrays. Arp220 shows maser emission from its West and East nuclei, with the
latter showing strong similarity to IIIZw35: Bright maser emission from two well
separated regions that contain about a half of the total source flux. 


The central velocities of the two bright regions differ by \about\ 100 \kms, yet
their spectral ranges overlap, just as in IIIZw35. This suggests that a similar
ring model might apply also to the Arp220 eastern nucleus; indeed, a rotation of
the model images presented in Figure \ref{fi:realizations1} shows a striking
similarity with the Arp220 maps (see Figure 2 in L98). The northern region of
the eastern nucleus contains more compact structures than the southern region,
including one very bright spot. This can be readily explained in our model,
which is stochastic in nature. Large velocity gradients over a short distance in
the bright regions are also explained by the overlap of clouds or filaments at
slightly different centre velocities. In the western nucleus of Arp 220 the
difference in compactness between the northern and southern patches of emission
are harder to reconcile with a rotating ring geometry, although the large LCRs
($>800$) and large $\rho$ ($>100$) are qualitatively explained by our multiple
cloud overlap model.

Another possible candidate source is Mrk273 \citep{K03}. This source, too, shows
both compact and resolved-out maser emission, and the compact emission occurs in
two distinct regions with different velocity centres. The existence of these two
regions can be explained by competition between the fall off of background
continuum brightness as we move away from the projected ring axis versus
increased path-length through the OH ring. The latter path-length increases the
maser brightness almost exponentially while the continuum strength effects it
only linearly.

Note that in IIIZw35, Mrk273 and Arp220 the brightest masers are not coincident
with the brightest continuum, but this is not strong evidence for saturated
emission as is sometimes claimed. Instead, given the exponential effect of path
length on brightness in unsaturated masers it is quite feasible that the
brightest emission occurs in regions where the continuum emission is very weak
or presently undetectable.

\section{Physical considerations\label{se:physics}}

\subsection{Cloud mass and confinement\label{se:physics_of_clouds}}

Here we discuss the physical properties of our OH maser clouds. The maximum
density of a OH maser emitting gas is of order $n(\rm H\sub{2})=\E{5}$ 
cm$^{-3}$. Higher densities will thermalise the energy levels and quench the
maser \citep{MOSHE92}. Assuming this maximum density, a OH abundance of \E{-5}
and excitation temperature 10K, equation 9.12 in \citet{MOSHE92} gives a minimum
cloud diameter of 0.02pc. Assuming instead that clouds have diameters equal to
the observational upper limit of 0.7pc then the same equation gives a minimum
mean hydrogen density of $3.1\x\E{3}$cm$^{-3}$. This gives an upper limit for
cloud mass of 24\Mo.  Note however that this  upper limit is  critically
dependant on the  assumed OH abundance.

Are the OH maser clouds confined? It seems that gravitational confinement can
be rejected. Given their internal velocity dispersion of 20\kms, to be
gravitationally bound the clouds would require a virial mass of \about
2\x\E{4}\Mo. This is much larger than the upper limit estimated from OH
properties. Such a cloud mass would also give a total mass in clouds of \about
2\x\E{7}\Mo, which is larger than the dynamical mass internal to the maser 
ring of 7\x\E{6}\Mo\ derived from the maser kinematics (P01).

Could the clouds be pressure confined by the ionised Inter-Cloud Medium (ICM)?
Using the inferred optical depth of the free-free absorbing gas and assuming it
is uniformly distributed, the electron number density in the ICM is \about\E{3}\
cm$^{-3}$. If the temperature of the ICM is \E{4} K, then the pressure becomes
$P\sub{ICM}\about\E{7}$ K cm$^{-3}$. In contrast, from the internal turbulent
velocity of the clouds their inferred dynamical pressure is of order 3\x\E{8}\ K
cm$^{-3}$, therefore they cannot be pressure confined. Finally magnetic
confinement would require a magnetic field of \about 10mG. This is feasible
given the range of magnetic field strengths observed towards galactic OH masers
(\citet{RE80}, \citet{FI89}) if the OH masers occurred in gas of the highest
possible density (i.e near $n(\rm H\sub{2})=\E{5}$ cm$^{-3}$). For comparison in
IIIZw35 Zeeman splitting observations by \citet{KI96} estimate an upper limit on
the non-uniform magnetic field to be 5mG, however this limit is very model
dependant.

Although magnetic confinement is possible given the observational limits, it is
interesting to consider models in which the clouds are freely expanding.
Assuming the largest possible cloud size (0.7pc) and dividing by the cloud
internal velocity dispersion gives a characteristic cloud lifetime of \about
3.4\x\E{4} yr. Given their outflow velocity of 60\kms\ in the $z$ direction,
such clouds would reach a height of \about 2.1 pc which is roughly consistent
with the observed $\Delta H/2$=3pc (see Table \ref{ta:table1}). An additional
advantage of this model is that that the ionised free-free absorbing gas
required by our model might naturally be generated from the dissipated clouds.
If the clouds have uniform density and consequently a mass of \about 24\Mo\ then
the total kinetic power being injected into all clouds would be about 7\x\E{38}\
erg s$^{-1}$ which is a very small fraction of the mechanical power available
from supernova explosions (i.e. 2\x\E{43}\ erg s$^{-1}$) assuming a supernova rate
of 0.8yr$^{-1}$ (P01). Using the same cloud mass,
the mass loss rate in the outflowing clouds is 0.8\Mo yr$^{-1}$ which is much
smaller than the estimated SFR (19\Mo yr$^{-1}$, see P01)

\subsection{Constraints on central AGN and Black Hole} \label{se:BH}

Our axisymmetric model (Figure \ref{fi:figure2}) suggests a central point to
both the ring and the ionised outflow and it is natural to ask whether an AGN 
or  black hole exists at this point. We should first remark that within the OH
ring radius, the geometry of our model is observationally ill-defined and so the
cone of free-free absorbing gas shown in figure \ref{fi:figure2} may not in fact
extend all the way inwards to a central point. If as we argue in
\S\ref{se:physics_of_clouds}, the free-free absorbing gas is the remains of
dissipated molecular clouds, then the free-free cone will in fact be truncated
at a height comparable to the OH ring height. 

Observationally there is conflicting evidence for the presence of an AGN in
IIIZw35. The source lies on the well known FIR-radio correlation for starbursts
(see P01) and so has no radio excess which might accompany a strong AGN. Also
high resolution radio observations (see P01) do not show any compact radio
features at our inferred centre. However such a radio core might be free-free
absorbed at 1.6GHz so it would be interesting to conduct higher frequency VLBI
observations to check this possibility. We have not been able to find any X-ray
observations which imply an AGN. 

In contrast to the above C90 argued that the near-IR colours of IIIZw35 were
consistent with it being a mixed starburst/AGN. Furthermore from their optical
spectroscopy C90 classified IIIZw35 as a borderline LINER/Seyfert2 nucleus based
on classical (semi-empirical) line ratio diagnostics. A similar analysis by
\cite{BSL98} using different optical spectroscopic observations but a similar
diagnostic scheme classified IIIZw35 as a LINER. Whether galaxies classified as
LINERs are primarily AGN or starburst powered is presently unclear. Using the
physically based line diagnostic scheme of \cite{KEW01} many LINERs are
reclassified as high metallicity starbursts \citep{COR03}. However, taking
the data for IIIZw35 from \cite{BSL98} and applying the \cite{KEW01} diagnostic
scheme we find that IIIZw35 still falls in the AGN region. The data is however 
consistent with a mixed AGN/starburst with the latter still providing most of
the luminosity in the diagnostic optical lines. As noted by \cite{COR03}.
estimating the relative bolometric AGN contribution in starbursts purely 
from optical  emission lines is very difficult because of the large optical 
obscuration in  these objects.

Is there any dynamical evidence for a central black hole? From the observed
maser dynamics  the derived enclosed mass within the OH maser ring radius of
22pc is 7\x\E{6}\Mo\ . The resulting mass density of 156\Mo\ pc$^{-3}$ can
easily be achieved by stars or gas within the nuclear region and so there is no
necessity for a central point mass. However a moderate mass black hole is not
ruled out either. The total enclosed mass in IIIZw35 could for instance be
contributed equally from a black hole like the one in our galaxy (3.5\x\E{6}\Mo\
, \citet{EG97}) and a distributed mass of stars. We note that even in the
extreme case that such a black hole is being maximally fed, then the maximum
luminosity if radiating at 10\% of the Eddington limit would be 2\x\E{10}\Lo\
which is over ten times smaller than the observed FIR luminosity. We conclude
that although optical spectroscopy shows that an AGN might be present in IIIZw35
it must be energetically insignificant compared to the starburst activity.

\subsection{Nature of the OH and continuum rings}

We argue in the previous section that the bulk of the bolometric emission  in
IIIZw35 is powered by starburst activity rather than an AGN. Certainly  for the
radio emission its consistency with the radio-FIR correlation, its observed
brightness temperature (P01) and its ring-like morphology argues for an origin 
in a starburst ring. Such a starburst ring can also  provide the IR photons   to
pump OH and therefore also explain the ring morphology of the  maser emission. 

A remarkable property of the OH ring is its relative narrowness (see
\S\ref{se:dimensions} and Table \ref{ta:table1}), which is only 12\% of the
radius. The fact that OH absorption is seen outside the ring and perhaps in its
central hole (P01) suggests that this geometry is not defined merely by OH
abundance. The most direct interpretation of the narrow maser ring is that it is
defined by the range of radii over which star formation is presently occurring.
Such ring can be identified as a scaled down version of the sub-kiloparsec
nuclear starburst rings known to exist in many starburst galaxies (see
\cite{KN04} and references therein). In addition, numerical simulations of
starburst-generated rings or tori have been made (\cite{WA03}, \cite{WA05})
supporting the existence of circumnuclear filamentary structures. The presence
of radio continuum emission at larger radii in IIIZw35 (see Figure
\ref{fi:figure1}) may indicate the remnant of past star formation, in turn
suggesting an inwardly propagating ring of star-formation. The fact that the OH
masing clouds in the ring are not gravitationally bound to the nucleus (see
\S\ref{se:physics_of_clouds}) suggests that emission at any given radius must be
short lived. Supernova explosions in a ring may compress dense gas on the inside
of the ring causing the ring to propagate inwards. 

Alternatively to the above picture, star formation in IIIZw35 may exist over a
wide range of radii, rather than just in a ring, but the required conditions for
population inversion might only apply over a narrow range of radii. The
inversion of OH main  lines is critically dependant on the local IR spectrum
(\cite{MOSHE78},  \cite{CO88}, \cite{MO88}, \cite{KE99}), which in turn depends
on the local SFR rate and Initial Mass Function.

It is interesting to compare the size and structure of our OH maser ring with
other OH maser observations of circumnuclear features. In Mrk231 \citep{K03} an
OH torus with radius from 30pc to 100pc has been inferred. It follows that the 
inner edge of that torus is comparable in size to the ring in IIIZw35 but it
extends over a larger radius and appears to be much thicker. We argue in
\S\ref{se:othersources} that the structures seen at the eastern nucleus of
Arp220 \citep{RO02} could also be explained by a thin ring with a radius of
\about 30pc being very similar to IIIZw35. It appears from the observations that
a wide variety of circumnulear OH maser structures can exist.  

The OH torus found in Mrk 231 \citep{K03} has been identified with the obscuring
torus required in AGN unified schemes, although the inner radii of such
structures are expected to be much smaller (close to the dust sublimation radius
which is sub-parsec in weak Seyferts or strong LINERS). In contrast it does not
seem that the OH ring we see in IIIZw35 can directly contribute much to any such
obscuration. From the direction of observation, the ring does not cover the
central point where any AGN might be expected to reside. Furthermore the narrow
ring covers only a small solid angle and even if it was observed edge-on, the
obscuration would still be small because given the derived cloud column
densities (\S\ref{se:physics_of_clouds}) the expected obscuration is only $A_{v}
\about 1$ assuming a standard gas to dust ratio. However, it cannot be ruled out
that the structure seen in OH is the outer part of a starburst supported disk
which extends down to small radii where it might have a much larger geometrical
and optical thickness.

\section{Conclusions and future work}
\label{se:conclusions}

Observations of distinct regions of compact and apparently diffuse maser
emission in OH megamasers have, in the past, been used to argue for two physical
phases of OH masers (see L98, D99). In contrast, the mechanism discussed in this
paper explains both types of maser structures using a single phase of low
opacity  clouds within a thin circumnuclear ring. The properties of the clouds
are similar to those assumed in the standard model \citep{BAAN89} but our model
explicitly considers the statistical effects of rare multi-cloud overlaps.
Compared to others, our model is therefore conservative in that the physical
properties of the standard model are preserved.

Despite the simplicity of our model we find that most of the observational
features of IIIZw35 can be reproduced. The fact that the maser amplifying medium
is composed of clouds is found essential to explain the range of maser
brightness around the ring at low resolution. The same clouds also explain the
bright maser spots seen only at the ring tangents in terms of multiple cloud
overlaps in both space and velocity. The model is able to reproduce the LCRs in
the bright spots and diffuse regions, and the large value of the 1667MHz:1665MHz
line ratio in the compact spots. Finally it can explain the spectra of both the
compact spots and the apparently diffuse areas of emission. The fact that the OH
clouds are outflowing from the ring midplane explains the large velocity
gradients observed amongst the compact maser spots. We find that the OH masing
ring is relatively narrow in radius which could be explained either in terms of
a narrow circumnuclear ring of star formation, or due to the strong sensitivity
of maser pumping to physical conditions which vary gradually with radius. The
ring we find is narrower but qualitatively similar to those that have been
produced in numerical simulations (\cite{WA02}, \cite{WA05}).

Our model can be improved in many ways. It assumes all clouds are identical,
when in reality a spectrum of cloud sizes and opacities is expected.
Additionally it presently assumes spherical clouds whereas computer simulations
(\cite{WA02}) show that filamentary structures are common in starburst nuclei.
These latter structures may also show outward velocity gradients due to
acceleration by radiation pressure; including such filaments will give rise to
higher velocity correlation between masers spots, as seen in the observations.
The next step in simulations of megamaser sources might be to take the output of
numerical dynamical simulations and calculate the expected OH megamaser
emission.

One line of future work involves taking the estimates of cloud opacity, size,
internal velocity dispersion and 1667:1665 MHz line ratio found in IIIZw35 and
deriving the implications for gas physical conditions and pumping. Another
important area to investigate is the general one of properties of maser emission
in a cloudy medium, especially the generalisation to media with a range of cloud
opacities.  While much work has been put into understanding masers in simple
geometries such as spheres or filaments relatively little work has been done for
random media. One exception is \cite{SO03} who carried out numerical simulations
of masers in continuous media with a Kolmogorov spectrum of local opacity.
Multi-phase media are common in astrophysics and the case we have studied here
of masers arising in discrete clouds is another limit which requires careful
attention. Observationally separate regions of compact and apparently diffuse
maser emission are often found in other species of galactic and circumstellar
masers. Examples include galactic methanol \citep{MI00} and circumstellar SiO
masers \citep{YI05}. These core-halo structures are often interpreted in terms
of single clouds which have saturated outer layers and unsaturated cores.
However, such structures might also be explained by invoking the presence of a
clumpy medium.

\appendix{
\section{Maser gain in clumpy media}

Consider a population  of identical unsaturated maser clouds amplifying
background continuum. At a given velocity let \Nbar\ be the mean number of
clouds which are encountered along a given LOS. If the clouds are of non-zero
size and cannot interpenetrate there is also a maximum number of clouds \Nmax,
set by the number than can be fitted along a LOS. If the maser optical depth of
each cloud is $\tau\sub{c}$ then the mean value of the gain is

\eq{\label{eq:series}
G\sub{mean} = \sum_{k=0}^{N\sub{max}} P_{\bar
N, N\sub{max} }(k)\cdot e^{k \tau\sub{c}}, }

\noindent where $P_{\bar N, N\sub{max}}(k)$ is the probability of having $k$
overlaps. If the cloud volume filling factor is small such that $\Nmax >> \Nbar$
then $P_{\bar N, N\sub{max}}(k)$ can be replaced by the Poisson distribution
$P_{\bar N}(k) = \bar N^{k} e^{-\bar N} / k!$ and the upper limit on the sum
tends to infinity. In this case the mean of the gain is given by 

\eq{\label{eq:NP84}
G\sub{mean} = \sum_{k=0}^{N_{\rm max} \rightarrow \infty} P_{\bar N}(k)\cdot
e^{k\tau\sub{c}} =\exp\left[\Nbar(e^{\tau\sub{c}}-1)\right]}

\noindent as found by \cite{NP84}. However we find that when $\tau\sub{c}>1$
the above expression enormously overestimates the average gains in our
Monte-Carlo simulations; even for a negligible cloud filling factor. The reason
is that in the above series although the Poisson probabilities $P_{\bar N}(k)$
decreases rapidly for $k > \Nbar$, for $\tau\sub{c}>1$  this decrease is more
than offset by  the increase of the amplification factors $e^{k \tau\sub{c}}$.
The result is that the series summation becomes dominated by very rare events
with $k \gg \Nbar$. In contrast for a typical Monte-Carlo simulation such an
extreme multi-cloud overlap event is very unlikely to occur even once within
any of the independent lines of sight within an interferometer beam.

An analytic expression for $G\sub{mean}$ which is in closer agreement with our
Monte-Carlo simulations can be constructed in the following way: If the
interferometer beam is $M$ times larger than a cloud area, each beam can be
assumed to contain $M$ independent lines of sight. Consequently, the series in
equation \ref{eq:NP84} can be truncated at a finite \Nmax\, to exclude
improbable events that are unlikely to occur even once within the beam. More
accurately, \Nmax\ can be defined such that the maximum number of overlaps
within a region of $M$ lines of sight is less than \Nmax\ in 50\% of
realisations. This gives:

\eq{\label{eq:truncated}
G\sub{mean} \simeq \sum_{k=0}^{N_{\rm max}} P_{\bar N}(k)\cdot e^{k\tau\sub{c}}
\qquad\quad N\sub{max} = P_{\bar N}^{-1}\left(2^{-\frac{1}{M}} \right)
}

\noindent where $P_{\bar N}^{-1}$ is the inverse Poisson cumulative
distribution. Using Monte-Carlo simulations (see Figure \ref{fi:app_poisson})
this equation is found to give a good approximation to the mean gain over a
wide range of \Nbar. Using the above definition of  \Nmax\ the median value of
the peak gain within a region of $M$ independent lines of sight is given by

\eq{\label{eq:nmax}
     G\sub{max} \simeq \exp( N\sub{max}\cdot\tau\sub{c} )
}

An important point, which is evident from Figure \ref{fi:app_poisson}, is that
the effective  opacity $ \tau\sub{eff} = \ln(G\sub{mean})$ shown on the scale
to the right, is much larger than the mean opacity in that region
$\tau\sub{mean} = \Nbar\tau\sub{c}$ (shown at the scale on the top). The two
quantities differ because of the high $k$ tail of the  $P\sub{\bar N}$
distribution. Figure \ref{fi:app_poisson} also demonstrates another remarkable
property of masers in a clumpy medium, that the effective opacity
$\tau\sub{eff}$, is not linearly dependent on \Nbar\ but falls well below that
prediction as \Nbar\ and $\tau\sub{mean}$ increase. This effect arises because
as \Nbar\ increases the variance about the mean in the number of clouds per line of 
sight decreases which  in turn reduces the contribution from rare 
many-cloud overlaps. As described in \S\ref{se:need4clumps}  this nonlinear 
relationship between effective opacity and \Nbar\ for a clumpy media  
allows us to understand the otherwise puzzling variation in LCR around 
the maser ring.

\begin{figure} \centering
\includegraphics[width=0.95\columnwidth]{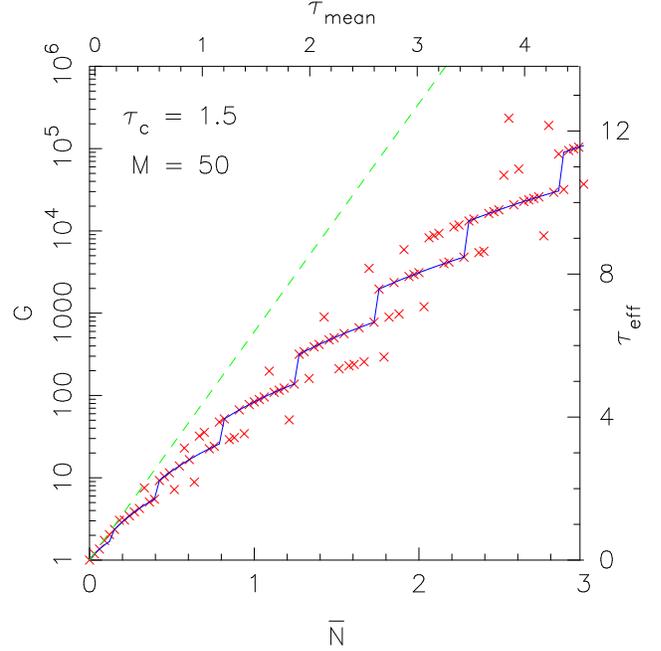}
\caption{Variation of the mean gain $G$ with the mean number \Nbar\ of
overlapping maser clouds when the individual cloud optical depth is
$\tau\sub{c}$ = 1.5. The crosses show the results of a numerical Monte-Carlo
calculation that averages over $M$ = 50 lines of sight. The solid line is the
result of analytic calculation using equation \ref{eq:truncated}. The dashed
line plots the result of an infinite-series summation (Equation \ref{eq:NP84}).
The top axis shows the distribution mean optical depth $\tau\sub{mean} =
\Nbar\tau\sub{c}$, the right axis its effective optical depth $\tau\sub{eff} =
\ln G\sub{mean}$.} \label{fi:app_poisson} \end{figure}

\begin{acknowledgements} R.P.\ thanks the University of Kentucky for a most
delightful month, which helped in bringing this paper forward to submission. J.C
gratefully acknowledges  support from  the Swedish science research council
(VR).  M.E.\ gratefully acknowledges the partial support of NSF. The authors
acknowledge the incisive comments sent by the referee,  Phil Diamond, which
helped to significantly improve the manuscript. \end{acknowledgements}


\bibliography{2971}

\begin{thebibliography}{38}
\expandafter\ifx\csname natexlab\endcsname\relax\def\natexlab#1{#1}\fi

\bibitem[{{Baan}(1989)}]{BAAN89}
{Baan}, W.~A. 1989, \apj, 338, 804

\bibitem[{{Baan} {et~al.}(1998){Baan}, {Salzer}, \& {Lewinter}}]{BSL98}
{Baan}, W.~A., {Salzer}, J.~J., \& {Lewinter}, R.~D. 1998, \apj, 509, 633

\bibitem[{{Chapman} {et~al.}(1990){Chapman}, {Staveley-Smith}, {Axon}, {Unger},
  {Cohen}, {Pedlar}, \& {Davies}}]{CH90}
{Chapman}, J.~M., {Staveley-Smith}, L., {Axon}, D.~J., {et~al.} 1990, \mnras,
  244, 281

\bibitem[{{Cohen} {et~al.}(1988){Cohen}, {Baart}, \& {Jonas}}]{CO88}
{Cohen}, R.~J., {Baart}, E.~E., \& {Jonas}, J.~L. 1988, \mnras, 231, 205

\bibitem[{{Corbett} {et~al.}(2003){Corbett}, {Kewley}, {Appleton},
  {Charmandaris}, {Dopita}, {Heisler}, {Norris}, {Zezas}, \& {Marston}}]{COR03}
{Corbett}, E.~A., {Kewley}, L., {Appleton}, P.~N., {et~al.} 2003, \apj, 583,
  670

\bibitem[{{Diamond} {et~al.}(1999){Diamond}, {Lonsdale}, {Lonsdale}, \&
  {Smith}}]{D99}
{Diamond}, P.~J., {Lonsdale}, C.~J., {Lonsdale}, C.~J., \& {Smith}, H.~E. 1999,
  \apj, 511, 178

\bibitem[{{Diamond} {et~al.}(1989){Diamond}, {Norris}, {Baan}, \&
  {Booth}}]{D89}
{Diamond}, P.~J., {Norris}, R.~P., {Baan}, W.~A., \& {Booth}, R.~S. 1989,
  \apjl, 340, L49

\bibitem[{{Eckart} \& {Genzel}(1997)}]{EG97}
{Eckart}, A. \& {Genzel}, R. 1997, Bulletin of the American Astronomical
  Society, 29, 1366

\bibitem[{{Elitzur}(1978)}]{MOSHE78}
{Elitzur}, M. 1978, \aap, 62, 305

\bibitem[{{Elitzur}(1992)}]{MOSHE92}
{Elitzur}, M. 1992, {Astronomical masers} (Astronomical masers Kluwer Academic
  Publishers (Astrophysics and Space Science Library.~Vol.~170), 365 p.)

\bibitem[{{Fiebig} \& {Guesten}(1989)}]{FI89}
{Fiebig}, D. \& {Guesten}, R. 1989, \aap, 214, 333

\bibitem[{{Heckman}(2003)}]{HE03}
{Heckman}, T.~M. 2003, in Revista Mexicana de Astronomia y Astrofisica
  Conference Series, 47--55

\bibitem[{{Henkel} \& {Wilson}(1990)}]{HW90}
{Henkel}, C. \& {Wilson}, T.~L. 1990, \aap, 229, 431

\bibitem[{{Kegel} {et~al.}(1999){Kegel}, {Hertenstein}, \&
  {Quirrenbach}}]{KE99}
{Kegel}, W.~H., {Hertenstein}, T., \& {Quirrenbach}, A. 1999, \aap, 351, 472

\bibitem[{{Kewley} {et~al.}(2001){Kewley}, {Heisler}, {Dopita}, \&
  {Lumsden}}]{KEW01}
{Kewley}, L.~J., {Heisler}, C.~A., {Dopita}, M.~A., \& {Lumsden}, S. 2001,
  \apjs, 132, 37

\bibitem[{{Killeen} {et~al.}(1996){Killeen}, {Staveley-Smith}, {Wilson}, \&
  {Sault}}]{KI96}
{Killeen}, N.~E.~B., {Staveley-Smith}, L., {Wilson}, W.~E., \& {Sault}, R.~J.
  1996, \mnras, 280, 1143

\bibitem[{{Kl{\" o}ckner} {et~al.}(2003){Kl{\" o}ckner}, {Baan}, \&
  {Garrett}}]{K03}
{Kl{\" o}ckner}, H., {Baan}, W.~A., \& {Garrett}, M.~A. 2003, \nat, 421, 821

\bibitem[{{Kl{\" o}ckner} \& {Baan}(2004)}]{K04}
{Kl{\" o}ckner}, H.-R. \& {Baan}, W.~A. 2004, \aap, 419, 887

\bibitem[{{Knapen} {et~al.}(2004){Knapen}, {Whyte}, {de Blok}, \& {van der
  Hulst}}]{KN04}
{Knapen}, J.~H., {Whyte}, L.~F., {de Blok}, W.~J.~G., \& {van der Hulst}, J.~M.
  2004, \aap, 423, 481

\bibitem[{{Lonsdale}(2002)}]{LSDL02}
{Lonsdale}, C.~J. 2002, in in Proceedings of the IAU Symposium 206, Cosmic
  Masers: From Proto-Stars to Black Holes, Ed. V. Mineese and M. Reid, San
  Francisco: Astronomical Society of the Pacific, 2002, Page 413

\bibitem[{{Lonsdale} {et~al.}(1994){Lonsdale}, {Diamond}, \& {Smith}}]{LSDL94}
{Lonsdale}, C.~J., {Diamond}, P.~J., \& {Smith}, H.~E. 1994, \nat, 370, 117

\bibitem[{{Lonsdale} {et~al.}(1998){Lonsdale}, {Lonsdale}, {Diamond}, \&
  {Smith}}]{LSDL98}
{Lonsdale}, C.~J., {Lonsdale}, C.~J., {Diamond}, P.~J., \& {Smith}, H.~E. 1998,
  \apjl, 493, L13+

\bibitem[{{Minier} {et~al.}(2000){Minier}, {Booth}, \& {Conway}}]{MI00}
{Minier}, V., {Booth}, R.~S., \& {Conway}, J.~E. 2000, \aap, 362, 1093

\bibitem[{{Mirabel} \& {Sanders}(1987)}]{MS87}
{Mirabel}, I.~F. \& {Sanders}, D.~B. 1987, \apj, 322, 688

\bibitem[{{Montgomery} \& {Cohen}(1992)}]{MC92}
{Montgomery}, A.~S. \& {Cohen}, R.~J. 1992, \mnras, 254, 23P

\bibitem[{{Moore} {et~al.}(1988){Moore}, {Mountain}, {Yamashita}, \&
  {Selby}}]{MO88}
{Moore}, T.~J.~T., {Mountain}, C.~M., {Yamashita}, T., \& {Selby}, M.~J. 1988,
  \mnras, 234, 95

\bibitem[{{Natta} \& {Panagia}(1984)}]{NP84}
{Natta}, A. \& {Panagia}, N. 1984, \apj, 287, 228

\bibitem[{{Pihlstr{\" o}m} {et~al.}(2001){Pihlstr{\" o}m}, {Conway}, {Booth},
  {Diamond}, \& {Polatidis}}]{P01}
{Pihlstr{\" o}m}, Y.~M., {Conway}, J.~E., {Booth}, R.~S., {Diamond}, P.~J., \&
  {Polatidis}, A.~G. 2001, \aap, 377, 413

\bibitem[{{Reid} \& {Moran}(1981)}]{RE80}
{Reid}, M.~J. \& {Moran}, J.~M. 1981, \araa, 19, 231

\bibitem[{{Rovilos} {et~al.}(2003){Rovilos}, {Diamond}, {Lonsdale}, {Lonsdale},
  \& {Smith}}]{RO02}
{Rovilos}, E., {Diamond}, P.~J., {Lonsdale}, C.~J., {Lonsdale}, C.~J., \&
  {Smith}, H.~E. 2003, \mnras, 342, 373

\bibitem[{{Smith} {et~al.}(1998){Smith}, {Lonsdale}, {Lonsdale}, \&
  {Diamond}}]{S98b}
{Smith}, H.~E., {Lonsdale}, C.~J., {Lonsdale}, C.~J., \& {Diamond}, P.~J. 1998,
  \apjl, 493, L17+

\bibitem[{{Sobolev} {et~al.}(2003){Sobolev}, {Watson}, \& {Okorokov}}]{SO03}
{Sobolev}, A.~M., {Watson}, W.~D., \& {Okorokov}, V.~A. 2003, \apj, 590, 333

\bibitem[{{Staveley-Smith} {et~al.}(1987){Staveley-Smith}, {Cohen}, {Chapman},
  {Pointon}, \& {Unger}}]{SS87}
{Staveley-Smith}, L., {Cohen}, R.~J., {Chapman}, J.~M., {Pointon}, L., \&
  {Unger}, S.~W. 1987, \mnras, 226, 689

\bibitem[{{Trotter} {et~al.}(1997){Trotter}, {Moran}, {Greenhill}, {Zheng}, \&
  {Gwinn}}]{T97}
{Trotter}, A.~S., {Moran}, J.~M., {Greenhill}, L.~J., {Zheng}, X., \& {Gwinn},
  C.~R. 1997, \apjl, 485, L79+

\bibitem[{{Wada} \& {Norman}(2002)}]{WA02}
{Wada}, K. \& {Norman}, C.~A. 2002, \apjl, 566, L21

\bibitem[{{Wada} \& {Norman}(2003)}]{WA03}
{Wada}, K. \& {Norman}, C.~A. 2003, in Astronomical Society of the Pacific
  Conference Series, 261--+

\bibitem[{{Wada} \& {Tomisaka}(2005)}]{WA05}
{Wada}, K. \& {Tomisaka}, K. 2005, \apj, 619, 93

\bibitem[{{Yi} {et~al.}(2005){Yi}, {Booth}, {Conway}, \& {Diamond}}]{YI05}
{Yi}, J., {Booth}, R.~S., {Conway}, J.~E., \& {Diamond}, P.~J. 2005, \aap, 432,
  531

\end{thebibliography}
\bibliographystyle{aa}

\end{document}